\documentclass[aps,prb,twocolumn,nofootinbib,superscriptaddress]{revtex4-1}% from Revtex sample

\usepackage[dvipdfmx]{graphicx}%picture
\usepackage{color}
\usepackage{comment}

\usepackage{amsmath, amssymb,bm}

\newcommand{\1}{\mbox{1}\hspace{-0.25em}\mbox{l}}

\begin{document}

%\preprint{}

\title{Chiral-symmetry protected exceptional torus in correlated nodal-line semi-metals}


\author{Kazuhiro Kimura}
\email[]{E-mail address: kimura.kazuhiro.85n@st.kyoto-u.ac.jp}
\affiliation{Department of Physics, Kyoto University, Kyoto 606-8502, Japan}

\author{Tsuneya Yoshida} %\\
\affiliation{Department of Physics, University of Tsukuba, Tsukuba, Ibaraki 305-8571, Japan}

\author{Norio Kawakami} %\\
\affiliation{Department of Physics, Kyoto University, Kyoto 606-8502, Japan}


\date{\today}

\begin{abstract}
We analyze a diamond-lattice Hubbard model with the spatially modulated Hubbard interaction. Our dynamical mean-field analysis with special emphasis on non-Hermitian properties elucidates that the gapless nodal line changes into symmetry-protected exceptional torus (SPET) at the Fermi level enclosing the three-dimensional open Fermi surface, which is unique to non-Hermitian physics with chiral symmetry. 
Furthermore, we also elucidate the effects of the SPETs on the magnetic response; our results based on the random-phase approximation combined with the dynamical mean-field theory shows that SPETs enhance the magnetic susceptibility at the weakly correlated sites, exemplifying effects of non-Hermitian degeneracies on responses to external fields.
\end{abstract}

%\keywords{}

\maketitle

\section{Introduction}
\label{introduction}
Since the theoretical discovery of topological insulators\cite{Hasan_Kane_RMP_2010,Qi_Zhang_RMP_2011}, the notion of topology has become much more ubiquitous in condensed matter physics, and has been further generalized to gapless phases
\cite{Armitage_Mele_Vishwanath_RMP_2018} such as Weyl semimetals and nodal-line semimetals (NLSMs), where the relation between gapless points/lines and band topology is well understood from the view point of underlying symmetry. In addition, the notion of topology has been extended to strongly correlated systems, where an interplay between topology and Coulomb interaction hosts some exotic phenomena
\cite{Hohenadler_2011,Yamaji_2011,Yu_Xie_2011,Yoshida_BHZ_2012,Tada_2012,Wu_2012,Yoshida_BHZ_2013}, such as topological Kondo insulators\cite{Dzero_TKI_2012,Dzero_RMP_2016}, topological Mott insulators\cite{Pesin_Balents_2010,Yoshida_TMI_2014,Yoshida_TMI_2016,Yoshida_BSPT_2016}, the change of topological classification\cite{Fidkowski_Kitaev_2010,Fidkowski_Kitaev_2011,Turner_2011,Ryu_2012,Yao_2013,Qi_2013,Lu_2012,Levin_2011,Fidkowski_2013,Gu_2014,Hsieh_2014,Wang_2014,Morimoto_2015,Isobe_2015,Yoshida_2015,You_2014,Yoshida_2017,Yoshida_2018}, etc.

On the other hand, more recent studies have revealed another kind of interesting topological phase in non-equilibrium systems, which is described by a non-Hermitian effective Hamiltonian
\cite{Hatano_Nelson_1996,Bender_Boettcher_1998,Fukui_Kawakami_1998,
Guo_prl_2009,Ruter_nat_2010,Regensburger_nat_2012,Esaki_prb_2011,Sato_prog_2012,
Zhen_2015,Lee_prl_2016,Ashida_pra_2016,Lee_prl_2016,San_2016,Ashida_2017,Gong_ZHE_2017,Hassan_prl_2017,
Shen_prl_2018,Kawabata_PTSC_2018,Gong_TPNH_2018,Yao_prl_2018,Yao_prl2_2018,Kunst_prl_2018,
Edvardsson_prb_2019,Kawabata_nat_2019,Kawabata_symandtopo_2019}. 
Furthermore, as proposed by Kozii and Fu, even in equilibrium systems, such non-Hermitian topological properties
\cite{Kozii_Fu_2017} can appear due to the life time effects originating from self-energy 
\cite{Papaj_2018,Shen_QO_2018,Yoshida_NH_2018,Mcclarty_2019,Bergholtz_2019}. 
For example, if we describe the energy spectrum of quasiparticles having the complex self-energy in terms of an effective Hamiltonian, non-Hermitian physics related to gapless defective points naturally shows up in strongly correlated systems. Such emergent defective points (or exceptional points) of non-Hermitian effective Hamiltonian give rise to an open Fermi surface
\cite{Gonzalez_PRB2017,Molina_PRL2018,Zhou_2018,Cerjan_exp_2018,Zyuzin_2018,Yong_2017,Cerjan_theory_2018,Carlstom_knott_2019,Carlstom_ribbon_2018,Wang_NLSM_2019,Yang_2019,Moors_2019,Okugawa_2019,Budich_Bergholtz_2019,Yoshida_ER_2019,Kawabata_Bessyo_2019,Yoshida_mech_2019} in the energy spectrum called a bulk Fermi arc. 
Furthermore, it has been elucidated that the symmetry enriches\cite{Okugawa_2019,Budich_Bergholtz_2019,Yoshida_ER_2019,Kawabata_Bessyo_2019,Yoshida_mech_2019} the possible shapes of exceptional points and the Fermi arcs. To date, various non-Hermitian topological semi-metals have been reported\cite{Cerjan_exp_2018,Zyuzin_2018,Yong_2017,Cerjan_theory_2018,Carlstom_knott_2019,Carlstom_ribbon_2018,Wang_NLSM_2019,Yang_2019,Moors_2019,Okugawa_2019} . For example, in two dimensions, the exceptional points form symmetry-protected exceptional rings, which induce Fermi planes. In three dimensions, they form symmetry-protected exceptional surfaces enclosing the region where the band gap becomes pure-imaginary.

In spite of the intensive studies, it remains unclear what physical properties are affected by the non-Hermitian band structure. 
In particular, there are few studies elucidating effects of the non-Hermitian band structure on magnetic/electric responses.

In this paper, we investigate emergent non-Hermitian properties in strongly correlated NLSMs with chiral symmetry, and discuss their impact on bulk quantities such as the magnetic susceptibility.
The chiral-symmetric NLSMs provide a feasible platform to study non-Hermiticity and symmetry-protected topological degeneracy.
Specifically, employing the dynamical mean-field theory\cite{Metzner_1989,Muller_1989,Georges_Kotliar_1992,Kajueter_Kotliar_1996,Georges_Kotliar_1996} 
combined with the iterated perturbation theory\cite{Georges_Kotliar_1992,Kajueter_Kotliar_1996,Georges_Kotliar_1996} 
(DMFT+IPT), we elucidate the emergence of symmetry-protected exceptional torus (SPETs) for a Hubbard model of the diamond lattice. 
These SPETs induce a sharp peak of the local density of states at the Fermi energy only for one of the sublattices having weak correlation, which results in the local magnetic susceptibility of strong sublattice dependence. To our best knowledge, this is the first result exemplifying how the non-Hermitian degeneracies affect magnetic responses.
We stress that the chiral symmetry is essential for the above behaviors. Recently, the emergence of SPETs with $PT$ (product of parity and time-reversal) symmetry has been reported by analyzing a noninteracting non-Hermitian Hamiltonian\cite{Okugawa_2019}. In contrast to such a case, SPETs with chiral symmetry are fixed to the Fermi level, which induces the Fermi volumes (i.e., low energy excitations enclosed by SPETs) in the Brillouin zone
\footnote{
SPETs with $CP$ (product of parity and particle-hole) symmetry are fixed to the Fermi level. However, the system with $CP$ symmetry is realized for superconductors.
}.

The rest of this paper is organized as follows. 
In Sec. \ref{model}, we describe our setup and give a brief explanation of our approach.
In Sec. \ref{result}, we study the emergence of exceptional torus at the Fermi level and its impact on bulk properties thorough the magnetic susceptibility. 
The last section is devoted to a brief summary.

%******************************************************************************
\section{Model and Method\label{model}}
\subsection{Model Hamiltonian}

We study the two-band Hubbard model with spatially modulated on-site Hubbard interactions on the diamond lattice;
\begin{eqnarray}
\hat{H}&=&\sum_{\langle i\alpha,j\alpha^{\prime}\rangle \sigma}t_{ij}\hat{c}^{\dagger}_{i\alpha\sigma}\hat{c}_{j\alpha^{\prime}\sigma}
+\sum_{i\alpha}U_{\alpha}(\hat{n}_{i\alpha\uparrow}-\frac{1}{2})(\hat{n}_{i\alpha\downarrow}-\frac{1}{2}), \nonumber \\
\label{eq:hami}
\end{eqnarray}
where $\hat{c}^{\dagger}_{i\alpha\sigma}(\hat{c}^{\dagger}_{i\alpha\sigma})$ creates (annihilates) a fermion at the $i$-th site of sublattice $\alpha(=A,B)$ with spin $\sigma$ and $\hat{n}_{i\alpha\sigma}=\hat{c}^{\dagger}_{i\alpha\sigma}\hat{c}_{i\alpha\sigma}$.
$t\in \mathbb{R}$ is a hopping parameter and $U_{\alpha} \in \mathbb{R}$ is an on-site interaction.
The first term of the above Hamiltonian describes hopping of fermions between neighboring sites in the diamond-lattice whose primitive vectors are $\bm{a}_i ,\, (i=1,2,3)$: $\bm{a}_1= \frac{a}{2}(0,1,1)$, $\bm{a}_2= \frac{a}{2}(1,0,1)$, $\bm{a}_3= \frac{a}{2}(1,1,0)$.
The noninteracting term denotes the NLSM, which is protected by the chiral (sublattice) symmetry.
Details are given in Appendix \ref{appendix:free_Hami}.
We note that the many-body Hamiltonian Eq. (\ref{eq:hami}) preserves the many-body chiral symmetry at half filling which is defined by Eqs. (\ref{eq:chiral_sym}) and (\ref{eq:chiral_op}).
We expect that our toy model can be realized for cold atoms because in such systems the spatially modulated interactions are fabricated by the optical Feshbach resonance
\cite{Yamazaki_2010,Clark_2015}.
As pointed out in Ref.~\onlinecite{Okugawa_2019}, $PT$ symmetry may induce SPETs for the NLSMs with gain and loss, 
which indicates the presence of SPETs for corresponding correlated systems.
We stress, however, that the crucial difference from the $PT$ symmetric case is that for our system with many-body chiral symmetry, the low energy excitations induced by the SPETs appear strictly at the Fermi level, which enhances the magnetic susceptibility.

%***************************************************************
\begin{figure}[h]
\begin{center}
\includegraphics[width=0.8\hsize]{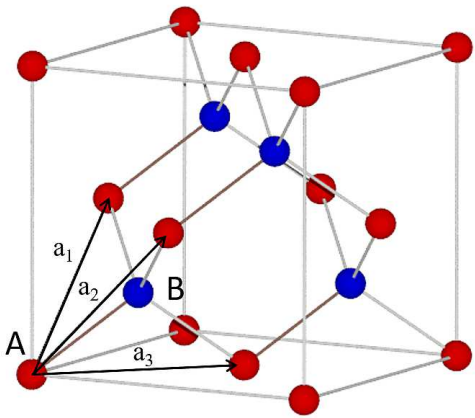}
\end{center}
\caption{
(Color Online) Sketch of the diamond lattice. Blue (red) spheres denote the $A$- ($B$-) sublattice.
$\bm{a}_i \, (i=1,2,3)$ denote primitive lattice vectors. 
}
\label{fig:lattice}
\end{figure}  
%***************************************************************
\subsection{DMFT+IPT method}
In this paper, we demonstrate the correlation-induced SPETs by spatially modulated interactions in three dimensional systems.
We employ the DMFT+IPT method to analyze correlation effects and clarify how SPETs affect low-energy properties.
In order to treat inhomogenity with the DMFT framework, we employ the sublattice method\cite{Georges_Kotliar_1996}.
In the DMFT framework, the lattice model is mapped to an effective impurity model described by
\begin{eqnarray} 
Z_{\rm eff}^{\alpha}&=&
 \int \mathcal{D}\bar{c}_{0\alpha\sigma}\mathcal{D}c_{0\alpha \sigma} e^{-S_{\rm eff}^{\alpha}},\\
S_{\rm eff}^{\alpha}&=&-\int^{\beta}_0 d\tau \int^{\beta}_0d\tau^{\prime} \sum_{\sigma}\bar{c}_{0\alpha\sigma}(\tau)\mathcal{G}_{\alpha\sigma}^{-1}(\tau-\tau^{\prime})c_{0\alpha\sigma}(\tau^{\prime}) \nonumber \\
&&+U_{\alpha} \int^{\beta}_0 d\tau (n_{0\alpha\uparrow}(\tau)-\frac{1}{2})(n_{0\alpha\downarrow}(\tau)-\frac{1}{2}) ,
\end{eqnarray}
where $\mathcal{G}_{\alpha\sigma}(\tau)$ is the noninteracting Green's function of the effective impurity model for the sublattice $\alpha$, imaginary time $\tau$, inverse temperature $\beta$ and $\bar{c}_{0\alpha\sigma}$ is a Grassmannian variable corresponding to the creation operator.
$\mathcal{G}_{\alpha\sigma}(\tau)$ is obtained by solving the following self-consistent equation:
\begin{eqnarray}
\mathcal{G}_{\alpha \sigma}^{-1}( \omega)&=&
\left[\frac{1}{N}\sum_{\bm{k}} \frac{1}{(\omega +i\delta +\mu)\1 -h(\bm{k})-\Sigma^R_{\sigma}(\omega)} \right]_{\alpha \alpha}^{-1}\nonumber \\
&&-\Sigma^R_{\alpha \sigma}(\omega),
\label{eq:self-eq}
\end{eqnarray}
where the Fourier representation of noninteracting Hamiltonian $h(\bm{k})$ is given by a $2\times 2$ matrix which is constructed from two sublattices and $\Sigma^R_{\sigma}(\omega):={\rm diag}(\Sigma^R_{A\sigma}(\omega), \Sigma^R_{B\sigma}(\omega))$ denotes the self-energy of the retarded Green's function describing electrons on sublattice $\alpha$.

In order to solve the self-consistent equation (\ref{eq:self-eq}), we employ the IPT method which is particularly efficient for a particle-hole symmetric system\cite{Georges_Kotliar_1996}.
The second-order self-energy is calculated as $\Sigma_{\alpha\sigma}^{(2)}(\tau)=-U_{\alpha}^2 \mathcal{G}_{\alpha\sigma}(\tau)\mathcal{G}_{\alpha-\sigma}(-\tau)\mathcal{G}_{\alpha-\sigma}(-\tau)$.
Here, we consider the spin symmetric case $\mathcal{G}_{\alpha \downarrow}=\mathcal{G}_{\alpha \uparrow}$.
Thus, the retarded self-energy is written as
\begin{eqnarray}
\Sigma_{\alpha}^{R(2)}(\omega)&=&
U^2\int^{\infty}_{-\infty} dx \int^{\infty}_{-\infty} dy \int^{\infty}_{-\infty} dz \rho^{0}_{\alpha}(x)\rho^{0}_{\alpha}(y)\rho^{0}_{\alpha}(z) \nonumber \\
&& \frac{f(-x)f(-y)f(z)+f(x)f(y)f(-z)}{\omega-x-y+z +i\delta},
\end{eqnarray}
where $\rho^{0}_{\alpha}(\omega)=-\frac{1}{\pi}{\rm Im}\Bigl[ {\rm tr}\mathcal{G}_{\alpha\sigma}(\omega + i\delta)\Bigr]$
is the density of state (DOS) and $\delta=0+$.
By using the IPT solver we can provide real-frequency self-energies and Green's function without a numerical analytic-continuation.

%%%%%%%%%%%%%%%%%%
\subsection{Physical Quantities}
%%%%%%%%%%%%%%%%%%
From obtained Green's function with DMFT+IPT, we compute the magnetic susceptibility as follows.
Based on the random-phase approximation (RPA), we obtain the site-resolved spin susceptibility 
$\chi_A^s:=(\chi_{AA}^{\rm RPA}+\chi_{AB}^{\rm RPA})/2$,
$\chi_B^s:=(\chi_{BB}^{\rm RPA}+\chi_{BA}^{\rm RPA})/2$ 
(the factor $1/2$ means the square of spin $1/2$ times spin degrees of freedom $2$) 
with $\bm{\chi}^{\rm RPA}(\bm{q},i\epsilon_m):=(\bm{1}-\bm{\chi}^0 \bm{U})^{-1}{\bm{\chi} ^0}$,
where 2$\times$2 matrices $\bm{\chi}^{\rm RPA}$ ($\bm{\chi} ^0$)  
are the RPA susceptibility (susceptibility with bubble approximation).
$\bm{U}$ denotes the interaction matrix, $\bm{U}:={\rm diag}(U_A,U_B)$.
Matrix elements $\chi_{\alpha \beta}^0$ are defined as
\begin{eqnarray}
\chi_{\alpha \beta}^0(\bm{q},i\epsilon_m)
&=&-\frac{T}{N}\sum_{\bm{k},n}
G_{\alpha \beta}(\bm{q}+\bm{k},i\omega_n+i\epsilon_m)
G_{ \beta \alpha}(\bm{k},i\omega_n),\nonumber \\
\label{eq:bubble}
\end{eqnarray} 
where $\epsilon_m=2m\pi T$, $m\in \mathbb{Z}$ and $G_{\alpha \beta}(\bm{k},i\omega_n)$ is lattice Green's function obtained from DMFT.
Here, we have used the relation $G_{\alpha}(\bm{k},i\omega_n)=\int_{-\infty}^{\infty}dx \frac{A_{\alpha}(\bm{k}, x)}{i\omega_n-x}$.
The magnetic susceptibility is given by the $\bm{q}=0$ and $\epsilon_m=0$ component of $\chi_{\alpha}^s(\bm{q},i\epsilon_m)$, $\alpha=A,B$.

%******************************************************************************
\section{Results\label{result}}
Firstly, we show that SPETs with chiral symmetry emerges for our model after a short review of symmetry-protected non-Hermitian degeneracies.
Secondly, based on the RPA approximation, we numerically elucidate that low energy excitations accompanying the SPETs enhance the magnetic susceptibility.

\subsection{Symmetry protection of exceptional torus for two band model with chiral symmetry\label{result:A}}

Let us analyze a generic system with chiral symmetry which has two bands.
In this case, the non-Hermitian effective Hamiltonian, describing the single-particle excitations, is defined as $H_{\rm eff}(\omega, \bm{k}):=h(\bm{k})+\Sigma^{R}(\omega +i\delta,\bm{k})$, where the Hermitian matrix $h(\bm{k})$ denotes the one-body part of the Hamiltonian, and $\Sigma^{R}(\omega +i\delta,\bm{k})$ denotes the self-energy with an infinitesimal positive constant $\delta$.
For a non-Hermitian 2$\times$2 Hamiltonian, we can write an effective Hamiltonian of the form:
\begin{eqnarray}
H_{\rm eff}=[b_0(\bm{k})+i d_0(\bm{k})]\tau_0+[\bm{b}(\bm{k})+i \bm{d}(\bm{k})]\cdot \bm{\tau},
\end{eqnarray}
with two real $d$-vectors $\bm{b}(\bm{k}):=(b_1(\bm{k}),b_2(\bm{k}),b_3(\bm{k}))$ and $\bm{d}(\bm{k}):=(d_1(\bm{k}), d_2(\bm{k}), d_3(\bm{k}))$ and real numbers $b_0(\bm{k}), d_0(\bm{k})$, where $\tau$'s are Pauli matrices.
We immediately find the eigenenergies of the form
\begin{eqnarray}
E_{\pm}(\bm{k})&=&b_0(\bm{k})+ id_0(\bm{k})\pm \sqrt{\bm{b}^2(\bm{k})-\bm{d} ^2(\bm{k})+ 2i \bm{b}(\bm{k})\cdot \bm{d}(\bm{k})}.\nonumber \\ 
\label{eq:energy_spectrum}
\end{eqnarray}
In order for the effective Hamiltonian to possess band touching points, two real $d$-vectors need to satisfy 
\begin{eqnarray}
\bm{b}^2(\bm{k})&=& \bm{d} ^2(\bm{k}), \, \,  \bm{b}(\bm{k})\cdot \bm{d}(\bm{k})=0.
\label{eq:defective}
\end{eqnarray}
If the solution has non-zero vectors, then these equations describe exceptional points where the theory becomes defective, i.e., the Hamiltonian cannot be diagonalized and lacks a complete basis of eigenvectors.

Now, let us analyze symmetry protection of the many-body chiral symmetry in strongly correlated systems. 
The definition of many-body chiral symmetry \cite{Gurarie_PRB2011,Manmana_PRB2012} for the many-body Hamiltonian $\hat{H}$ is:
\begin{eqnarray}
\hat{U}^{\dagger}_{\Gamma}\hat{H}^{*}\hat{U}_{\Gamma}&=&\hat{H},
\label{eq:chiral_sym}
\end{eqnarray}
where $\hat{U}_{\Gamma}$ is a unitary operator which transforms a creation and annihilation operator (see Appendix \ref{appendix:definition}).
The explicit form of the operator $\hat{U}_{\Gamma}$ is defined for systems composed of two sublattices, as follows:
\begin{eqnarray}
\hat{U}_{\Gamma}&=&\prod_{j s}(\hat{c}^{\dagger}_{js\uparrow}+{\rm sgn}(s)\hat{c}_{js\uparrow})(\hat{c}^{\dagger}_{js\downarrow}+{\rm sgn}(s)\hat{c}_{js\downarrow}),
\label{eq:chiral_op}
\end{eqnarray}
where ${\rm sgn}(s)$ takes $1$ and $-1$ for $s=A$ and $s=B$, respectively. 
In terms of Green's function (see Appendix \ref{appendix:derivation}), we obtain the following relation: 
\begin{eqnarray}
G(\omega +i\delta)&=&-U^{\dagger}_{\Gamma}G^{\dagger}{(-\omega +i\delta)}U_{\Gamma}, 
\end{eqnarray}
where $U_{\Gamma}$ is the chiral matrix and $G(\omega +i \delta)$ is the single-particle Green's function which is represented by $H_{\rm eff}$: $G^{-1}(\omega +i \delta)=\omega \1-H_{\rm eff}(\omega ,\bm{k})$.
The many-body chiral symmetry\cite{Yoshida_ER_2019} results in the following constraint on the non-Hermitian effective Hamiltonian: 
\begin{eqnarray}
H_{\rm eff}(\omega, \bm{k})&=&-U^{\dagger}_{\Gamma}H_{\rm eff}^{\dagger}(-\omega, \bm{k})U_{\Gamma}.
\label{eq:heff_symm}
\end{eqnarray}

In particular, at $\omega=0$, this constraint is reduced to $H_{\rm eff}(0, \bm{k})=-U^{\dagger}_{\Gamma}H_{\rm eff}^{\dagger}(0, \bm{k})U_{\Gamma}$, which we refer to as extended chiral symmetry.
In our model, the chiral matrix is written as $U_{\Gamma}:=\tau_3$.
From this constraint, each term of the effective Hamiltonian is divided into symmetric or anti-symmetric sectors as, 
\begin{eqnarray}
U_{\Gamma}^{\dagger} [b_i\tau_i ]^{\dagger} U_{\Gamma}
&=&
\left\{
\begin{array}{ll}
+b_i \tau_i, &\quad (i=0,3) \\
-b_i \tau_i, &\quad (i=1,2)
\end{array}
\right.,
\\
U_{\Gamma}^{\dagger} [i d_i\tau_i ]^{\dagger} U_{\Gamma}&=&
\left\{
\begin{array}{ll}
-i d_i\tau_i , &\quad (i=0,3) \\
i d_i\tau_i , &\quad (i=1,2)
\end{array}
\right.,
\label{eq:pauli_matrix}
\end{eqnarray}
where four parameters $b_i (i=1,2)$ and $d_i (i=0,3)$ respect the chiral symmetry. 
We note $b_0=b_3=d_1=d_2=0$.
Finally, considering the many-body chiral symmetry, the effective Hamiltonian $H_{\rm eff}(0, \bm{k})$ is expanded by Pauli matrices $\tau$'s as follows,
\begin{eqnarray}
H_{\rm eff}(0, \bm{k})&=&b_1(\bm{k}) \tau_1+ b_2(\bm{k}) \tau_2  +i \{ d_0(\bm{k}) \tau_0+d_3(\bm{k}) \tau_3\}. \nonumber \\ 
\label{eq:heff}
\end{eqnarray}
Thus, the second condition of Eq. (\ref{eq:defective}) is satisfied automatically by the chiral symmetry.
The manifold consisting of defective points is determined by the single constraint $\bm{b}^2(\bm{k})= \bm{d} ^2(\bm{k})$.
The number of conditions for band degeneracy is reduced to one.
As a result, a $(d-1)$-dimensional exceptional manifold emerges in the $d$-dimensional system when $d \geq 0$ \cite{Yoshida_ER_2019}. 
Thus, we can obtain an exceptional ring and torus in two- and three- dimensional systems with chiral symmetry.
Note that the Fermi surface emerges in the region surrounded by exceptional manifold, called Femi volumes\cite{Budich_Bergholtz_2019}, which are open regions of vanishing real part of the energy gap and have the same dimension as the system itself.
%******************************************************************************

%***************************************************************
\begin{figure}[h]
\begin{center}
\includegraphics[width=\hsize]{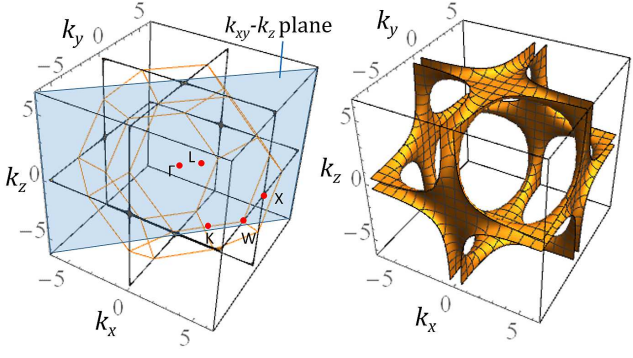}
\end{center}
\caption{
(Color Online).
Left panel: the Fermi surface (the black line) at the noninteracting case in the three-dimensional BZ.
Right panel: the SPET (the orange surface) for $(U_A/t,U_B/t)=(8,0)$ and $T/t=0.8$.
In the left figure, the orange line (blue plane) represents the three-dimensional BZ ($k_{xy}-k_z$ plane).
}
\label{fig:def}
\end{figure}  
%***************************************************************

%***************************************************************
\begin{figure}[h]
\begin{center}
\includegraphics[width=\hsize]{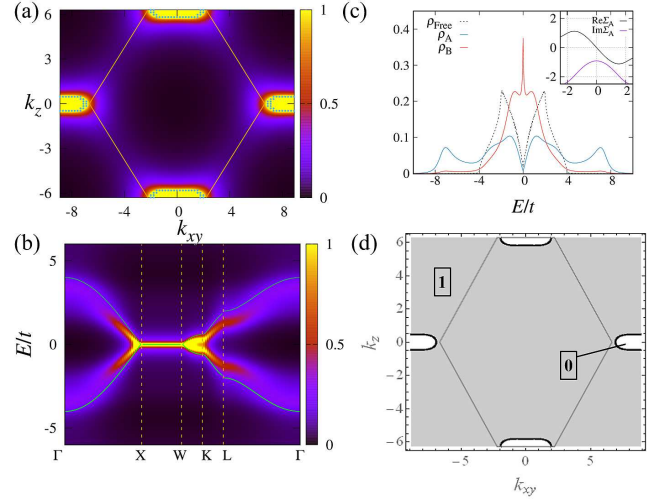}
\end{center}
\caption{
(Color Online)
(a) [(b)]: Momentum-resolved spectral weight $A(\bm{k},\omega=0)$ at $k_{xy}$-$k_{z}$ plane [$A(\bm{k},\omega)$ at high symmetric points] of NLSM with $(U_A/t,U_B/t)=(8,0)$ at temperature $T/t=0.8$. In panel (a), orange lines illustrate the BZ, the effective Hamiltonian becomes defective at the blue dots and the color plot indicates the strength of spectral weight.
In panel (b), the blue line denotes the energy spectrum of free diamond lattice.
(c): Density of states for each sublattice for $(U_A/t,U_B/t)=(8,0)$ at temperature $T/t=0.8$.
(d): Color map of the zero-th Chern number for ${\rm Im}\Sigma^{R}_A(\omega=0)=-0.92$ at $k_{xy}$-$k_{z}$ plane.
Here, the zero-th Chern number at each point is indicated by a number enclosed with a box and SPETs are represented with black line.
}
\label{fig:fig3}
\end{figure}  
%***************************************************************
\subsection{DMFT results}
Having finished general arguments on chiral symmetric SPETs in the previous section, we now present the DMFT results in this section. 
We first obtain the effective Hamiltonian from DMFT+IPT calculation. 
The effective Hamiltonian $H_{\rm{eff}}(0,\bm{k})$ is expanded in terms of the Pauli matrices $\tau$'s as follows:$H_{\rm{eff}}(0,\bm{k})=id_0(\bm{k})\tau_0+[\bm{b}(\bm{k})+i\bm{d}(\bm{k})]\cdot \bm{\tau}$ with $\bm{b}(\bm{k}):=(b_1(\bm{k}),b_2(\bm{k}),0)$ and $\bm{d}(\bm{k}):=(0,0,d_3(\bm{k}))$, which are given by
\begin{eqnarray}
\label{eq:fcomp}
b_1(\bm{k})+i b_2(\bm{k})&=&t_0+\sum_{j=1,2,3}t_j e^{i\bm{k}\cdot\bm{a}_j}, \\
d_0(\bm{k})\tau_0+d_3(\bm{k})\tau_3&=&{\rm Im}\Sigma^{R}(0+i\delta ,\bm{k}). 
\label{eq:parameter}
\end{eqnarray}
Here, the Pauli matrices $\tau$'s act on the sublattice space and $t_i$ with $i=0,1,2,3$ denotes the nearest neighbor hopping indicated by silver (gold) bonds in Fig.\ref{fig:lattice}.
In this case, by using the above effective Hamiltonian, the Green's function is written as 
\begin{eqnarray}
G(\omega, \bm{k})^{-1}&=&(\omega +i\delta)\tau_0 - H_{\rm{eff}}(\omega,\bm{k}).
\end{eqnarray}

\subsubsection{Interaction-driven SPET}
First, we show the noninteracting Fermi surface. In the left panel of Fig.\ref{fig:def}, we can confirm that the diamond lattice shows nodal-line semi-metals.
Introducing on the interaction, the nodal line changes to the SPET (see left panel of Fig. \ref{fig:def}). 
Inside of the SPET, the gap becomes pure imaginary and low energy excitations appear as a Fermi volume. 
We discuss more details about the non-Hermitian band structure in Appendix \ref{appendix:NHband}. 
The emergence of Fermi volume which is surrounded by SPET can be seen via the momentum-resolved spectral function $A(\bm{k}, \omega)=-\frac{1}{\pi}{\rm Im}\Bigl[ {\rm tr}G^{R}(\bm{k},\omega)\Bigr]$ [Fig. \ref{fig:fig3}(a)]. 
Away from the Fermi level, these low energy excitations are smoothly connected with the renormalized bands see [Fig. \ref{fig:fig3}(b)]. 
Here is a brief comment on Fig. \ref{fig:fig3}(b). Although Fig. \ref{fig:fig3}(b) seems to suggest that the region of high spectral weight between the ${\rm X}$ and ${\rm W}$ points has finite width in energy, we have confirmed that this width hardly depends on the interaction strength. 
The emergent Fermi volume, which is formed mainly by the contribution of the noninteracting $B$-sublattice, 
gives rise to a peak structure in the LDOS for the $B$-sublattice at zero energy as seen in Fig. \ref{fig:fig3}(c). Note that these characteristic features come from largely different lifetimes for the two sublattices.
The peak structure of LDOS is not limited to the diamond lattice but generally applies for two-sublattice chiral symmetric system.
We discuss the relationship between the origin of the peak structure and the many-body chiral symmetric system in Appendix \ref{appendix:ldos}.
The appearance of SPET and the peak structure of LDOS originate from the sublattice-dependent Hubbard interaction.

%***************************************************************
\begin{figure}[h]
\begin{center}
\includegraphics[width=\hsize]{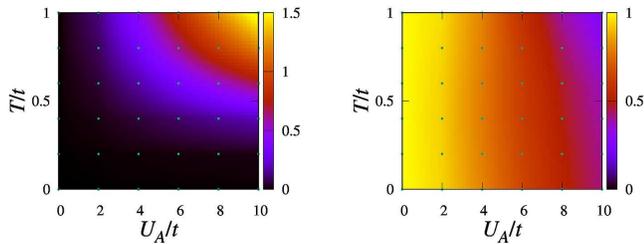}
\end{center}
\caption{(Color Online) 
Left panel: Color plot of the imaginary part of the self-energy $-{\rm Im}\Sigma^R_A(\omega=0)$ as functions of temperature $T/t$ and interaction $U_A/t$.
Right panel: Color plot of the renormalization factor  $z_{\alpha}=[1-\partial {\rm Re}\Sigma^{R}_{\alpha}(\omega)/\partial \omega]^{-1}$ as functions of temperature $T/t$ and interaction $U_A/t$.
These data are obtained for the isotropic diamond lattice at $U_B/t =0$.
We note that the imaginary part of the self-energy determines the radius of SPETs; the radius is $r=-{\rm Im}\Sigma^R_A(\omega=0)/2$.
}
\label{fig:phase}
\end{figure}  
%***************************************************************

Second, we check topological properties of the system.
We show that a zero-th Chern number can be introduced for the occupation number of the following Hermitian Hamiltonian\cite{Yoshida_ER_2019,Kawabata_Bessyo_2019} $\tilde{H}(\bm{k}):=-i H'(\bm{k}) \Gamma$ with $H'(\bm{k}):=H_{\rm eff}(0,\bm{k})-[{\rm tr}H_{\rm eff}(0,\bm{k})/N]\1$ and $N:={\rm dim}H_{\rm eff}(0,\bm{k})$.
Exceptional points are consistent with the change of the Chern number in Fig.\ref{fig:fig3}(d).
The radius of SPETs is related  to the difference in the imaginary part of self-energy $r=[{\rm Im}\Sigma^{R}_B(\omega)-{\rm Im}\Sigma^{R}_A(\omega)$]/2 at $\omega=0$ between $A$- and $B$- sublattices (see inset of Fig.\ref{fig:fig3}(c)).
We show the size of SPET as functions of $A$-site interaction $U_A/t$ and temperature $T/t$ in Fig.\ref{fig:phase}. We confirm that the radius of SPET is enhanced with increasing $A$-site interaction or temperature.
However the renormalization factor of the $A$-site Green's function hardly depends on temperature in Fig.\ref{fig:phase}.

\begin{figure}[h]
\begin{center}
\includegraphics[width=\hsize]{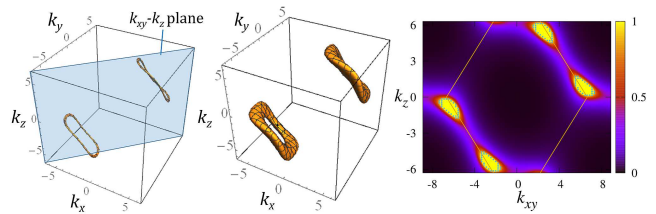}
\end{center}
\caption{
(Color Online) The left (middle) panel: Noninteracting Fermi surface of NLSM (ET with $(U_A/t,U_B/t)=(12,0)$ at temperature $T/t=1$) in the case of anisotropic diamond lattice $t_0/t_i =2, (i=1,2,3)$.
The right panel: Momentum-resolved spectral weight $A(\bm{k},\omega=0)$ on $k_z$-$k_{xy}$ plane with $(U_A/t,U_B/t)=(12,0)$ at temperature $T/t=1$. The orange lines illustrate the BZ at $k_{xy}$-$k_{z}$ plane and the effective Hamiltonian becomes defective at the blue dots in the right panel. 
}
\label{fig:aniso}
\end{figure}

We proceed to investigate the anisotropic diamond lattice with $t_0/t_i\neq1, i=1,2,3, t_i=t$.
With increasing the parameter of hopping with $1<t_0/t_i<3$, the topological phase transition occurs and we obtain various shapes of SPETs in Appendix \ref{appendix:aniso}.
We show the Fermi surface of nodal line semimetals with $t_0/t_i=2$ in Fig.\ref{fig:aniso} in the noninteracting case.
Considering spatially modulated on-site Coulomb interactions, we obtain SPETs and momentum-resolved spectral weight in Fig.\ref{fig:aniso} with $(U_A/t,U_B/t)=(12,0)$ at temperature $T/t=1$.

We stress that in contrast to the $PT$ symmetric case\cite{Okugawa_2019}, the band touching region inside of the SPETs emerges at $\omega=0$ (in this case it is called the Fermi volume) because of the chiral symmetry. 
This is understood by the prohibition of real coefficients of Pauli matrix $\tau_0$ in an effective Hamiltonian although the $PT$ symmetric system does not prohibit it.
As discussed momentarily below, the Fermi volume inside of SPETs affects low-energy physical properties of the system.

%***************************************************************
%***************************************************************

\subsubsection{Effects of SPETs on the magnetic susceptibility}
We show that the SPETs induce a nontrivial response to the magnetic field; the low energy excitations enclosed by the SPETs make the magnetic susceptibility of the $B$-sublattice larger than that of the $A$-sublattice, although the interaction strength is opposite ($U_A>U_B$).
The mechanisms are as follows. As seen in Fig.\ref{fig:fig3}(c), at $\omega=0$, the LDOS for $B$-sublattice becomes larger than that for $A$-sublattice because of SPETs (see Appendix \ref{appendix:ldos}). This fact indicates that the response to the Zeeman splitting for $B$-sublattice becomes larger than that for $A$-sublattice. Therefore, the magnetic moment of $B$-sublattice can becomes larger than that of $A$-sublattice.

\begin{figure}[h]
\begin{center}
\includegraphics[width=\hsize]{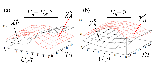}
\end{center}
\caption{
(Color Online).
The magnetic susceptibility for each sublattice $\chi_{\alpha}^s(\bm{q}=0,i\epsilon_m=0)$, $\alpha=A,B$ as functions of temperature $T/t$ and interaction $U_A/t$.
Panel (a) [(b)] are obtained at $U_B=U_A/2$ (at $U_B=0$) with RPA.
}
\label{fig:sus}
\end{figure}

The numerical data supporting the above scenario are shown in Fig.~\ref{fig:sus}(a) and (b). The former (the latter) data are obtained at $U_B=U_A/2$ ($U_B=0$). In Fig.~\ref{fig:sus}(a), we can see that the magnetic susceptibility for $B$-sublattice becomes larger than that for $A$-sublattice, corresponding to the emergence of the low energy excitations enclosed by the SPETs (for the LDOS at $U_B=U_A/2$, see Appendix \ref{appendix:ldos} and for the self-energy, see Appendix \ref{appendix:nonzeroUb}).
We can also observe similar behaviors at $U_B=0$ although the region of $\chi^s_B > \chi^s_A $ is narrow in this extreme limit.
We stress that essential ingredients are the difference in the lifetime of the self-energy and the chiral symmetry. Therefore, the above enhancement is generic and considered to be observed for any two-sublattice model where the self-energy satisfies $|{\rm Im} \Sigma^{R}_A(\omega=0)|> |{\rm Im} \Sigma^{R}_B(\omega=0)|$.
%***************************************************************
%***************************************************************

\section{Summary\label{summary}}
We have studied chiral-symmetric correlated NLSMs in three dimensions with special emphasis on non-Hermitian properties.
Concretely, we have elucidated the emergence of SPETs for a diamond lattice model with spatially modulated Hubbard interaction.
Essential difference from the case for NLSMs with $PT$ symmetry studied so far is that the present SPETs with chiral symmetry and the associated low energy excitations (i.e., Fermi volumes) are fixed to the Fermi level. 
Furthermore, we have elucidated that these low energy excitations result in counterintuitive behaviors which are the first results exemplifying the effects of SPETs on magnetic responses. 
Specifically, by employing by DMFT and RPA, we have found that due to the Fermi volumes, magnetic susceptibility for $B$-sublattice becomes larger than that for $A$-sublattice, although the interaction strength is opposite ($U_A>U_B$). 
For this counterintuitive response to the homogeneous magnetic field, the chiral symmetry is essential which leads to the enhancement of LDOS at the Fermi level only for $B$-sublattice.

One may wonder whether SPETs persist when we go beyond the DMFT-IPT\cite{Georges_Kotliar_1992,Kajueter_Kotliar_1996,Georges_Kotliar_1996}. 
We consider that they persist because of the following reasons. 
(i) It is well-known that the IPT method gives qualitatively correct results for systems with particle-hole symmetry. Indeed, the DMFT combined with the numerical renormalization group method elucidates the emergence of symmetry-protected exceptional rings for two-dimensional systems with chiral symmetry\cite{Yoshida_ER_2019} which can be regarded as a cross section of SPETs in three dimensions. 
(ii) The chiral symmetry ensures that the spatial fluctuations do not destroy the SPETs. The symmetry constraint in Eq. (\ref{eq:heff_symm}) for $\omega =0$ ensures that the effective Hamiltonian can be written in the form of Eq. (\ref{eq:heff}), indicating that the SPETs persist even in the presence of spatial fluctuations. The details study on this point is left for future work.

We finish this paper with comments on future studies.
Important open questions are (i) finding experimental setups or candidate materials showing SPETs and (ii) elucidating how to observe the unique magnetic response elucidated in this paper. 
Concerning the toy model analyzed in this paper, we expect that it can be realized for cold atoms by employing optical Feshbach resonance\cite{Yamazaki_2010,Clark_2015}. 
In such systems, the unique magnetic response might be observed by noise correlations or Bragg scattering of light which have been employed to observe the spin correlation functions\cite{Hart_2015,Mazurenko_2017}.
In addition, elucidating effects of non-Hermitian band structures on magnetic responses for other cases of symmetry remains an important future work.

\begin{acknowledgments}
This work was partly supported by JSPS KAKENHI Grant No. JP15H05855, JP18H01140, JP18H05842 and JP19H01838.
The numerical calculations were performed on the supercomputer at the Institute for Solid State Physics in the University of Tokyo, and SR16000 at Yukawa Institute for Theoretical Physics in Kyoto University.
\end{acknowledgments}
%******************************************************************************

\appendix

\section{Non-interacting Hamiltonian \label{appendix:free_Hami}}
\begin{figure}[h]
\begin{minipage}{\hsize}
\begin{center}
\includegraphics[width=\hsize]{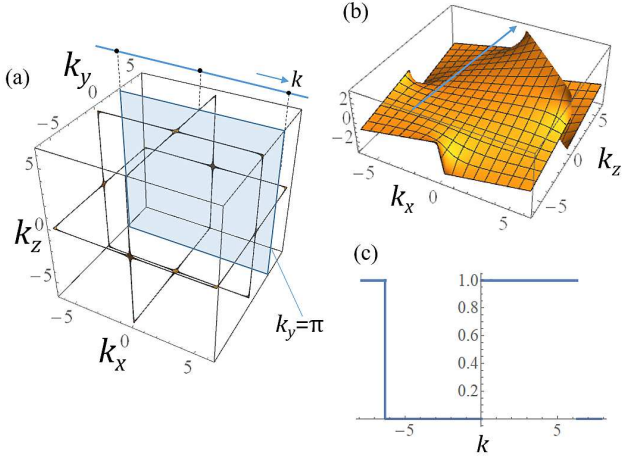}
\end{center}
\end{minipage}
\caption{
(Color Online). 
(a): The Fermi surface (the black line) at the noninteracting case in the three-dimensional BZ.
(b): The argument of the off-diagonal element $D_{\bm{k}}$ on $k_y=\pi$ plane and the blue arrow means the path of integral.
(c): The winding number $\nu_{k_x,k_y=\pi}$ on the blue arrow of (a).
In panel (b), we plotted the ${\rm Arg} \, D_{\bm{k}}$, where ${\rm Arg}$ means the principal value of the argument within the interval ($-\pi,\pi$].
}
\label{fig:fig7}
\end{figure}

The nodal-line structure of the noninteracting Hamiltonian can be understood by stacking of the one-dimensional topological insulators with chiral symmetry.  
The noninteracting Hamiltonian $h(\bm{k})$ has the chiral symmetry: $\{h(\bm{k}), \tau_3 \} =0$, where $\tau_3$ is the chiral matrix which acts on the sublattice degrees of freedom in two sublattice system.
The noninteracting Hamiltonian $h(\bm{k})$ reads,
\begin{eqnarray}
h(\bm{k})&=&
\begin{pmatrix}
0 & D_{\bm{k}} \\
D^{*}_{\bm{k}}&0
\end{pmatrix}
, \, \, 
D_{\bm{k}}=t_0+\sum_{j=1,2,3}t_j e^{i\bm{k}\cdot\bm{a}_j}.
\end{eqnarray}
To see the topological properties of the Dirac line node, we calculate the winding number along the blue line in Fig. \ref{fig:fig7}(a).
The definition of the winding number is:
\begin{eqnarray}
\nu_{k_x,k_y}&=&\frac{1}{2 \pi i}\int_{-2\pi}^{2\pi} dk_z \partial_{ k_z} {\rm ln}  D_{\bm{k}},  \nonumber \\
&=&\frac{1}{2 \pi}[ {\rm arg} D_{\bm{k}} |_{k_z=2\pi}-{\rm arg} D_{\bm{k}}|_{k_z=-2\pi}].
\end{eqnarray}
We note that the period in the $k_z$ direction is $4\pi$.
The structure of the argument of $D_{\bm{k}}$ on $k_y=\pi$ plane is shown in Fig. \ref{fig:fig7}(b) and the winding number is shown in Fig. \ref{fig:fig7}(c).
The change of winding number signals the bulk gapless structure and the Dirac line node structure appears in the three-dimensional BZ.

%***************************************************************
%***************************************************************

\section{Symmetry argument about the many-body chiral symmetry \label{appendix:many-body_chiral}}
\subsection{The definition of the many-body chiral symmetry\label{appendix:definition}}
First, the definition of many-body chiral symmetry \cite{Gurarie_PRB2011,Manmana_PRB2012} is 
\begin{eqnarray}
\hat{U}^{\dagger}_{\Gamma}\hat{H}^{*}\hat{U}_{\Gamma}&=&\hat{H},
\end{eqnarray}
where $\hat{H}$ is many-body Hamiltonian and $\hat{U}_{\Gamma}$ is the chiral operator which is a unitary operator $\hat{U}_{\Gamma}^2=1$. $\hat{U}_{\Gamma}$ transforms a creation and annihilation operator $\hat{c}^{\dagger}_{in}$ and $\hat{c}_{in}$ (where $i$ labels the sites of a lattice and $n$ labels the internal degrees of freedom such as sublattice and spin, etc), as follows: $\hat{U}^{\dagger}_{\Gamma}\hat{c}_{in}^{\dagger}\hat{U}_{\Gamma}=\sum_m  U_{\Gamma, nm}\hat{c}_{im}$ 
and $\hat{U}^{\dagger}_{\Gamma}\hat{c}_{in}\hat{U}_{\Gamma}=\sum_m \hat{c}_{im}^{\dagger}  U^{\dagger}_{\Gamma, mn}$, where the unitary matrix $U^{(\dagger)}_{\Gamma}$ is the chiral matrix of the noninteracting Hamiltonian $h_{ij}$ satisfying $U^{\dagger}_{\Gamma}h_{ij}U_{\Gamma}=-h_{ij}$ with $U_{\Gamma}^2=\1$. Here $\1$ denotes the identity matrix.
The explicit form of the chiral operator $\hat{U}_{\Gamma}$ is defined for systems composed of two sublattices, as follows:
\begin{eqnarray}
\hat{U}_{\Gamma}&=&\prod_{j s}(\hat{c}^{\dagger}_{js\uparrow}+{\rm sgn}(s)\hat{c}_{js\uparrow})(\hat{c}^{\dagger}_{js\downarrow}+{\rm sgn}(s)\hat{c}_{js\downarrow}),
\end{eqnarray}
where ${\rm sgn}(s)$ takes $1$ and $-1$ for $s=A$ and $s=B$, respectively. 
From the above chiral transformation, it is straightforward to see that: 
\begin{eqnarray}
\hat{U}^{\dagger}_{\Gamma}\hat{c}_{i s \sigma}^{\dagger}\hat{U}_{\Gamma}&=&{\rm sgn}(s)\hat{c}_{is \sigma},  \, \, 
\hat{U}^{\dagger}_{\Gamma}\hat{c}_{i s \sigma}\hat{U}_{\Gamma}={\rm sgn}(s)\hat{c}^{\dagger}_{is \sigma}. 
\end{eqnarray}
We note that the chiral matrix $U_{\Gamma}=\tau_3$, where the Pauli matrix $\tau_3$ acts on the sublattice space.

\subsection{Derivation of Eq. (\ref{eq:heff_symm})\label{appendix:derivation}}
We denote the many-body chiral symmetry in terms of Green's function\cite{Gurarie_PRB2011,Manmana_PRB2012,Yoshida_ER_2019} in Eq. (\ref{eq:heff_symm}).
We introduce the reterded and advanced Green's functions ($G^{R}(t)$, $G^{A}(t)$):
\begin{eqnarray}
G^{R}_{ab}(t)&=&-i\theta(t) [
\langle  \hat{c}_a(t)\hat{c}_b^{\dagger} (0) \rangle
+\langle  \hat{c}_b^{\dagger} (0)\hat{c}_a(t) \rangle
],\\
G^{A}_{ab}(t)&=&i\theta(-t) [
\langle \hat{c}_a(t)\hat{c}_b^{\dagger} (0) \rangle
+\langle  \hat{c}_b^{\dagger} (0)\hat{c}_a(t) \rangle
],
\end{eqnarray}
where $a$ and $b$ denote the set of indices, lattice site $i$ and internal degrees of freedom $n$; $\hat{c}_{a}:=\hat{c}_{in}$ and $\theta(t)$ is a step function.
Now we assume that $t>0$. Then we obtain the following relation:
\begin{eqnarray}
&&\langle  \hat{c}_a(t)\hat{c}_b^{\dagger} (0) \rangle   \nonumber \\
&=&{\rm Tr}[ e^{-\beta \hat{H}} e^{i\hat{H}t} \hat{c}_a(0)  e^{-i\hat{H}t} \hat{c}_b^{\dagger} (0)]  \nonumber \\
&=&{\rm Tr}[ e^{-\beta \hat{H}}  \hat{U}_{\Gamma}^{\dagger} e^{i\hat{H}^{*}t} \hat{U}_{\Gamma}\hat{c}_a(0)  \hat{U}_{\Gamma}^{\dagger}e^{-i\hat{H}^{*}t}\hat{U}_{\Gamma} \hat{c}_b^{\dagger} (0)\hat{U}_{\Gamma}^{\dagger}\hat{U}_{\Gamma} ],  \nonumber \\
&=&U_{\Gamma, aa^{\prime}}^{\dagger}U_{\Gamma, b^{\prime}b}
{\rm Tr}[ e^{-\beta \hat{H}^{*}} e^{i\hat{H}^{*}t} \hat{c}_{a^{\prime}}^{\dagger}(0)  e^{-i\hat{H}^{*}t} \hat{c}_{b^{\prime}}(0)],  \nonumber \\
&=&U_{\Gamma, aa^{\prime}}^{\dagger}U_{\Gamma, b^{\prime}b}
{\rm Tr}[ \hat{c}_{b^{\prime}}^{\dagger}(0)e^{-i\hat{H}t} \hat{c}_{a^{\prime}}(0) e^{i\hat{H}t}e^{-\beta \hat{H}} ],  \nonumber \\
&=& U_{\Gamma, aa^{\prime}}^{\dagger}U_{\Gamma, b^{\prime}b} 
\langle  \hat{c}_{b^{\prime}}^{\dagger} (0)\hat{c}_{a^{\prime}}(-t) \rangle.
\label{eq:time_green}
\end{eqnarray}
Here, we have used the following relations:
\begin{eqnarray}
e^{i\hat{H}t}&=&\hat{U}_{\Gamma}^{\dagger} e^{i\hat{H}^{*}t} \hat{U}_{\Gamma}, \, \, 
e^{-\beta \hat{H}}=\hat{U}_{\Gamma}^{\dagger} e^{-\beta \hat{H}^{*}} \hat{U}_{\Gamma},\\
\hat{U}_{\Gamma} \hat{c}_a \hat{U}_{\Gamma}^{\dagger}&=& U^{\dagger}_{\Gamma, aa^{\prime}}\hat{c}^{\dagger}_{a^{\prime}}, \, \,
\hat{U}_{\Gamma} \hat{c}_b^{\dagger} \hat{U}_{\Gamma}^{\dagger}= U^{\dagger}_{\Gamma, b^{\prime}b}\hat{c}_{b^{\prime}},\\
\langle M^*|\hat{A}|N^* \rangle&=&\langle N|\hat{A}^T| M\rangle,
\label{eq:cc}
\end{eqnarray}
where $|M\rangle$ and $|N\rangle$ denote general eigenstates of the many-body Hamiltonian. 
In a similar way, we have 
\begin{eqnarray}
\langle  \hat{c}_b^{\dagger}(0)\hat{c}_a (t) \rangle &=&
U_{\Gamma, aa^{\prime}}^{\dagger}U_{\Gamma, b^{\prime}b} 
\langle  \hat{c}_{a^{\prime}}(-t) \hat{c}_{b^{\prime}}^{\dagger} (0) \rangle.
\end{eqnarray}
As a result, we obtain $G^{R}(t)=-U^{\dagger}_{\Gamma}G^A(-t)U_{\Gamma}$ and the Fourier representation is:  
\begin{eqnarray}
G(\omega +i\delta)&=&-U^{\dagger}_{\Gamma}G{(-\omega -i\delta)}U_{\Gamma}, 
\label{eq:green_many-body_1}
\end{eqnarray}
where $G(\omega +i \delta)$ is Green's function.

To go from Eq. (\ref{eq:green_many-body_1}) to Eq. (\ref{eq:green_many-body_2}), we have used the following relation:
\begin{eqnarray}
G^{\dagger}_{ab}{(\omega +i\delta)}&=&G_{ab}(\omega -i\delta). 
\label{eq:green_dagger}
\end{eqnarray}
This relation is understood by the Lehmann representation of the Green's function, as follows:
\begin{eqnarray}
G_{ab}{(z)}&=&\sum_{NM} e^{\beta (\Omega -E_N)} \frac{ e^{\beta (E_N-E_M)}+1}{z+E_N-E_M} 
\langle N|\hat{c}_a|M\rangle \langle M|\hat{c}_b^{\dagger}|N\rangle, \nonumber \\
\end{eqnarray}
where $z \in \mathbb{C}$, $e^{\beta \Omega}=\sum_N e^{-\beta E_N}$, and $E_N ( \in \mathbb{R})$ is the energy for the many-body Hamiltonian.
So the Hermitian conjugate of the Green's function is obtained:
\begin{eqnarray}
G_{ab}^{\dagger}{(z)}&=&\sum_{NM} e^{\beta (\Omega -E_N)} \frac{ e^{\beta (E_N-E_M)}+1}{(z+E_N-E_M)^*} 
\langle N|\hat{c}_b|M\rangle ^*  \langle M|\hat{c}_a^{\dagger}|N\rangle  ^*, \nonumber \\
&=&\sum_{NM} e^{\beta (\Omega -E_N)} \frac{ e^{\beta (E_N-E_M)}+1}{z^*+E_N-E_M} 
\langle N|\hat{c}_a|M\rangle   \langle M|\hat{c}_b^{\dagger}|N\rangle , \nonumber \\
&=&G_{ab}{(z)},
\end{eqnarray}
where we have used Eq. (\ref{eq:time_green}) to go from the first line to the second line.
We thus obtain Eq. (\ref{eq:green_dagger}).

Combining Eq. (\ref{eq:green_dagger}) and Eq. (\ref{eq:green_many-body_1}), 
we arrive at the many-body chiral symmetry in terms of Green's function:
\begin{eqnarray}
G(\omega +i\delta)&=&-U^{\dagger}_{\Gamma}G^{\dagger}{(\omega -i\delta)}U_{\Gamma}. 
\label{eq:green_many-body_2}
\end{eqnarray}

The effective Hamiltonian $H_{\rm eff}(\omega ,\bm{k})$ is defined by the single-particle Green's function: $G^{-1}(\omega +i \delta)=\omega \1 -h_{\bm{k}}-\Sigma(\omega+i\delta ,\bm{k})=\omega \1-H_{\rm eff}(\omega ,\bm{k})$.
In terms of $H_{\rm eff}(\omega, \bm{k})$, we can write the constraint of the many-body chiral symmetry of the form:
\begin{eqnarray}
H_{\rm eff}(\omega, \bm{k})&=&-U^{\dagger}_{\Gamma}H_{\rm eff}^{\dagger}(-\omega, \bm{k})U_{\Gamma}.
\end{eqnarray}

\section{Non-Hermitian band structure \label{appendix:NHband}}
\begin{figure}[h]
\begin{center}
\includegraphics[width=0.9\hsize]{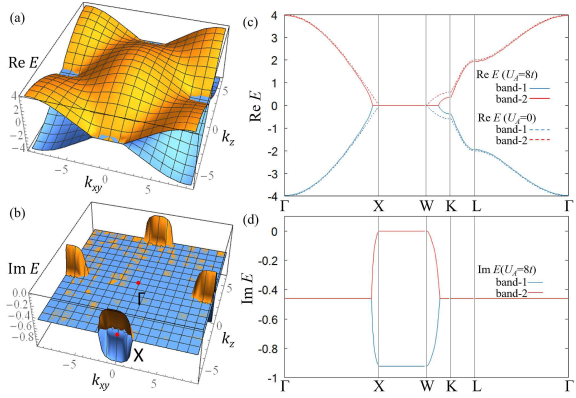}
\end{center}
\caption{
(Color Online).
The energy dispersion of non-Hermitian Hamiltonian $H_{\rm eff}(0,\bm{k})=h(\bm{k})+\Sigma (0)$ with $(U_A/t,U_B/t, T/t)=(8,0,0.8)$.
(a) and (b) [(c) and (d)] are the real and imaginary part of $H_{\rm eff}(0,\bm{k})$ on the $k_{xy}-k_z$ plane [high symmetric line]. 
In panel (c), solid (dotted) lines denote the dispersion with $U_A=8t$ ($U_A=0$).
}
\label{fig:fig8}
\end{figure}

We here discuss the energy dispersion of the non-Hermitian effective Hamiltonian $H_{\rm eff}(0,\bm{k})=h(\bm{k})+\Sigma (0)$.
The energy dispersion is shown in Fig. \ref{fig:fig8}(a)- \ref{fig:fig8}(d).
On the $k_{xy}-k_z$ plane, the real part of $H_{\rm eff}(0,\bm{k})$ in Fig. \ref{fig:fig8}(a) has the band touching region which corresponds to the Fermi surface in Fig. \ref{fig:fig3}(a) and the imaginary part in Fig. \ref{fig:fig8}(b) is gaped in this region.
The defective points in Fig. \ref{fig:fig8}(a) correspond to the band touching region where the real and imaginary part of $H_{\rm eff}(0,\bm{k})$ have simultaneously the gapless structure.
Next, we show the energy dispersion on a high symmetric line in Fig. \ref{fig:fig8}(c) and (d). 
The noninteracting band structure changes and the band touching region induced by the non-Hermiticity appears in Fig. \ref{fig:fig8}(c).

%***************************************************************
%***************************************************************

\section{LDOS structure of spin U(1) chiral symmetric system\label{appendix:ldos}}

%***************************************************************
\begin{figure}[h]
\begin{center}
\includegraphics[width=0.7\hsize]{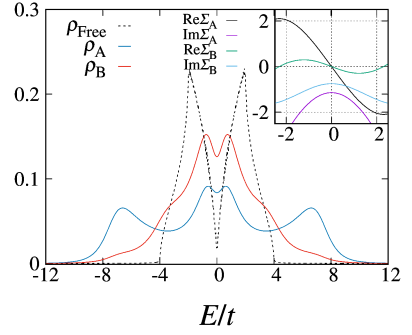}
\end{center}
\caption{
(Color Online).
Density of states for each sublattice for $(U_A/t,U_B/t)=(9,4.5)$ at temperature $T/t=0.5$.
The inset shows the self-energy $\Sigma^R_A(\omega)$ and $\Sigma^R_B(\omega)$ as a function of $\omega$.
}
\label{fig:fig9}
\end{figure}  
%***************************************************************

We show the origin of the peak structure in the LDOS for the $B$-sublattice at zero energy.
In general, we consider the two-sublattice system with many-body chiral symmetry\cite{Yoshida_ER_2019} and spin U(1) symmetry.
The full Green's function is written in a $2\times 2$ matrix which is constructed from two sublattices, as follows,
\begin{eqnarray}
G(\bm{k},\omega)&=&[\omega \1-h(\bm{k})-\Sigma^{R}(\bm{k},\omega)]^{-1},\\
&=&
\begin{pmatrix}
\omega-\Sigma^{R}_A & -D_{\bm{k}} \\
 -D^{*}_{\bm{k}}&\omega-\Sigma^{R}_B
\end{pmatrix}
^{-1}, 
\end{eqnarray}
where $h(\bm{k})$ and  $\Sigma^{R}(\bm{k},\omega)$ are Hamiltonian and self-energy,
\begin{eqnarray}
h(\bm{k})&=&
\begin{pmatrix}
0 & D_{\bm{k}} \\
D^{*}_{\bm{k}}&0
\end{pmatrix}
,
\Sigma^{R}(\bm{k},\omega)=
\begin{pmatrix}
\Sigma^{R}_A & 0 \\
0&\Sigma^{R}_B
\end{pmatrix}
.
\end{eqnarray}
$D_{\bm{k}}$ is the Fourier component of tight-binding model, for example Eq. (\ref{eq:fcomp}) and $\Sigma^{R}_{\alpha}$ are self-energy of each sublattice, $\alpha=A,B$.
We obtain full Green's functions of each subalttice, 
\begin{eqnarray}
G_{AA}(\bm{k},\omega)&=&\frac{\omega-\Sigma^{R}_B}{(\omega-\Sigma^{R}_A)(\omega-\Sigma^{R}_B)-|D_{\bm{k}}|^2},\\
G_{BB}(\bm{k},\omega)&=&\frac{\omega-\Sigma^{R}_A}{(\omega-\Sigma^{R}_A)(\omega-\Sigma^{R}_B)-|D_{\bm{k}}|^2}.
\end{eqnarray}
To discuss the structure of LDOS, we consider the $\omega=0$ component of spectral function for each sublattice, 
\begin{eqnarray}
A_{A}(\bm{k},\omega=0)&=&-\frac{1}{\pi}{\rm Im}G_{AA}(\bm{k},\omega=0),\\
&=&\frac{1}{\pi}\frac{-{\rm Im}\Sigma^{R}_{B}(0)}{{\rm Im}\Sigma^{R}_{A}(0){\rm Im}\Sigma^{R}_{B}(0)+|D_{\bm{k}}|^2},\nonumber \\ \\
A_{B}(\bm{k},\omega=0)&=&-\frac{1}{\pi}{\rm Im}G_{BB}(\bm{k},\omega=0),\\
&=&\frac{1}{\pi}\frac{-{\rm Im}\Sigma^{R}_{A}(0)}{{\rm Im}\Sigma^{R}_{A}(0){\rm Im}\Sigma^{R}_{B}(0)+|D_{\bm{k}}|^2},\nonumber \\
\end{eqnarray}
where $A_{\alpha}(\bm{k},\omega)$ is the spectral function of $\alpha$ sublattice, $\alpha=A,B$.
Now we consider the imbalance of self-energy at the Fermi level such as 
\begin{eqnarray}
-{\rm Im}\Sigma_{A}^{R}(\omega=0)&>&-{\rm Im}\Sigma_{B}^{R}(\omega=0),\\
|{\rm Im}\Sigma_{A}^{R}(\omega=0)|&>&|{\rm Im}\Sigma_{B}^{R}(\omega=0)|,
\label{eq:imbalance}
\end{eqnarray}
where, due to the positivity of the spectral function, ${\rm Im}\Sigma_{\alpha}(\omega=0)$ is always negative, $\alpha=A,B$.
The relation between spectral functions of two sublattices is obtained as,
\begin{eqnarray}
A_{A}(\bm{k},\omega=0)&<&A_{B}(\bm{k},\omega=0).
\end{eqnarray}
The above inequality is valid for all wave vectors in BZ. 
Finally, the LDOS is obtained from the summation of spectral function for all wave vectors 
and we show the relationship between LDOS of each sublattice, as follows,
\begin{eqnarray}
\rho_{A}(\omega=0)&<&\rho_{B}(\omega=0),
\end{eqnarray}
where $\rho_{\alpha}=\sum_{\bm{k}\in BZ}A_{\alpha}(\bm{k},\omega=0)$, $\alpha=A,B$.
The inequality means the imbalance of LDOS for each sublattice [see Figs.~\ref{fig:fig3}(c)~and~\ref{fig:fig9}].
In our case, the origin of self-energy is only on-site Coulomb interaction and the relation for the imbalance of self-energy Eq.(\ref{eq:imbalance}) is the same as the imbalance of on-site Hubbard interactions $U_A>U_B$ at finite temperature.
We show examples of the imbalance of self-energy in the inset of Figs.~\ref{fig:fig3}(c)~and~\ref{fig:fig9}.
As a result, in the many-body chiral symmetric system with spin U(1) symmetry, considering the spatially modulated on-site Hubbard interactions, the peak structure or the imbalance of LDOS structure always appears at on the Fermi level at finite temperatures.

\section{SPET of the anisotropic diamond lattice\label{appendix:aniso}}
\begin{figure}[h]
\begin{minipage}{\hsize}
\begin{center}
\includegraphics[width=\hsize]{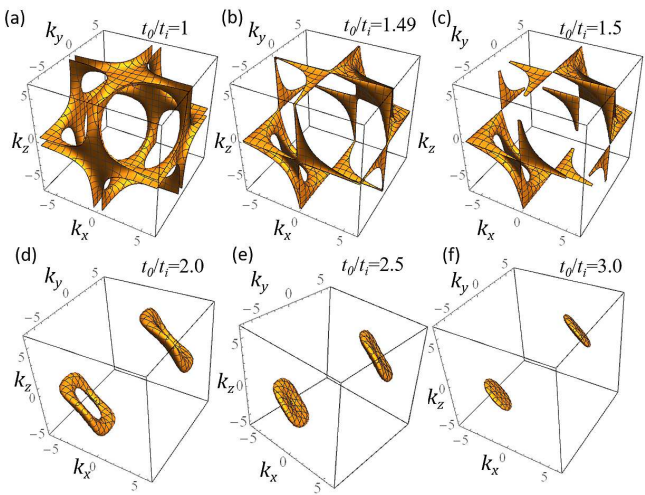}
\end{center}
\end{minipage}
\caption{
(Color Online). 
SPETs for anisotropic diamond lattices with $t_0/t_i\neq1, i=1,2,3, t_i=1$ and ${\rm Im}\Sigma_A^R(\omega=0)=-1$ in three-dimensional BZ.
}
\label{fig:fig10}
\end{figure}

We show the results of anisotropic diamond lattices in Fig.(\ref{fig:fig10}). 
The shape of SPETs is determined from the band touching point of Eqs. (\ref{eq:energy_spectrum}) and (\ref{eq:parameter}).
The topological phase transition occurs between $t_0/t_i=1.49$ and $t_0/t_i=1.5$, between $t_0/t_i=2.5$ and $t_0/t_i=3.0$.  
%***************************************************************
%***************************************************************

\section{SPET of isotropic diamond lattice for $U_B=U_A/2$\label{appendix:nonzeroUb}}
\begin{figure}[h]
\begin{minipage}{0.8\hsize}
\begin{center}
\includegraphics[width=\hsize]{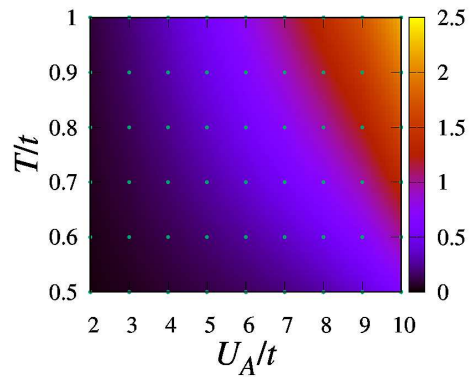}
\end{center}
\end{minipage}
\caption{ 
(Color Online).
Color plot of the difference in the imaginary part of the self-energy ${\rm Im}\Sigma^{R}_B(\omega=0)-{\rm Im}\Sigma^R_A(\omega=0)$ as functions of temperature $T/t$ and interaction $U_A/t$ for the isotropic diamond lattice at $U_B =U_A/2$.The imaginary part of the self-energy determines the radius of SPETs.
We note that the imaginary part of the self-energy determines the radius of SPETs; the radius is $r=[{\rm Im}\Sigma^{R}_B(\omega=0)-{\rm Im}\Sigma^R_A(\omega=0)]/2$.
}
\label{fig:fig11}
\end{figure}

We investigate the isotropic diamond lattice for $U_B=U_A/2$.
The obtained difference in the imaginary part of self-energy ${\rm Im}\Sigma_B^{R}(\omega=0)-{\rm Im}\Sigma^{R}_A(\omega=0)$ in Fig.~(\ref{fig:fig11}).
%***************************************************************
%***************************************************************
\newpage

%\bibliography{main}

%

\end{document}